\documentclass[sigconf]{acmart}

\usepackage[english]{babel}
\usepackage{blindtext}
\usepackage{subfigure}
\usepackage{url}

\renewcommand\footnotetextcopyrightpermission[1]{} % removes footnote with conference info
\setcopyright{none}
\acmDOI{}

\acmISBN{}

\acmPrice{}

\begin{document}
\title{V-CNN: When Convolutional Neural Network encounters Data Visualization}

%\titlenote{Produces the permission block, and copyright information}
%\subtitle{Extended Abstract}

\author{Mao Yang, Bo Li, Guanxiong Feng, Zhongjiang Yan}
 \affiliation{%
   \institution{Northwestern Polytechnical University, Xi'an, China}
   %\streetaddress{West Youyi Road NO. 127, Xi'an, China}
   %\city{Xi'an}
   %\state{China}
   %\postcode{710072}
 }
 \email{{yangmao, libo.npu, zhjyan}@nwpu.edu.cn}

% \author{Firstname Lastname}
% \authornote{Note}
% \affiliation{%
%   \institution{Affiliation}
%   \streetaddress{Address}
%   \city{City}
%   \state{State}
%   \postcode{Zipcode}
% }
% \email{email@domain.com}

\renewcommand{\shortauthors}{Mao Yang, et al.}

\begin{abstract}
In recent years, deep learning poses a deep technical revolution in almost every field and attracts great attentions from industry and academia. Especially, the convolutional neural network (CNN), one  representative model of deep learning, achieves great successes in computer vision and natural language processing. However, simply or blindly applying CNN to the other fields results in lower training effects or makes it quite difficult to adjust the model parameters. In this poster, we propose a general methodology named V-CNN by introducing data visualizing for CNN. V-CNN introduces a data visualization model prior to CNN modeling to make sure the data after processing is fit for the features of images as well as CNN modeling. We apply V-CNN to the network intrusion detection problem based on a famous practical dataset: AWID. Simulation results confirm V-CNN significantly outperforms other studies and the recall rate of each invasion category is more than 99.8\%.
\end{abstract}

\maketitle

\section{Introduction}
%\Blindtext

In recent years, deep learning becomes a driven force to deeply affect science and technology and is supposed to be an important enabler for achieving a new era. The performance of many fields have been significantly improved by using deep learning. Therefore, both industry and academia pay great attentions to deep learning, and several models or algorithms have been proposed.

The convolutional neural network (CNN) is one of the most successful and representative models of deep learning. Different from other artificial neural networks, CNN adopts the local area, named receptive field, information to construct the feed-forward neural network by introducing multiple convolutional layers and pooling layers, which means full connectivity of neurons is not necessary. CNN improves the convergence time and reduces the system complexity. In recent years, CNN achieves great success in computer vision and natural language processing because of the local information features. However, simply or blindly applying CNN to other fields, such as networking, always results in lower training performances or makes it quite difficult to adjust the model parameters. It means we encounter a paradox: the performance of CNN is quite attractive but the dataset of many other fields can hardly match CNN's features, which restricts the scope of CNN.

In this poster, to break the paradox and further improve the scope of CNN, we propose a general methodology named V-CNN by introducing data visualizing for CNN. V-CNN introduces a data visualization model prior to CNN model to make sure the original arbitrary data after processing is fit for the features of images as well as CNN model. After designing the architecture of V-CNN, we verify the performance by applying V-CNN to the network intrusion detection problem based on a famous dataset: AWID. Simulation results confirm that V-CNN significantly outperforms other studies and the recall rate of each invasion category is more than 99.8\%.

\section{Architecture of V-CNN}
%\blindtext
%\begin{figure}[tp]
%\centering
%\includegraphics{figures/mouse}
%\caption{\blindtext}
%\end{figure}
\begin{table*}[ht]
\centering
\caption{Results}\label{tab:aStrangeTable}%添加标题 设置标签
\begin{tabular}{cccccc}
\hline
Class& Records Number & \cite{ref_1} & \cite{ref_2} & \cite{ref_3} & V-CNN\\
\hline
Flooding & 16,682 & 99.94\% & 100.00\% & 99.94\% & \textbf{99.99\%}\\
Impersonation & 8,097 & 68.68\% & 73.78\% & 61.34\% & \textbf{100.00\%}\\
Injection & 20,079 & 99.75\% & 99.99\% & 82.72\% & \textbf{99.84\%}\\
Normal & 530,785 & 7.14\% & 22.00\% & 98.49\% & \textbf{99.95\%}\\
How many algorithms are used to \\achieve the best classification result? & ------ & 1 & 3 & 2 & \textbf{Only 1}\\
\hline
\end{tabular}
\end{table*}

\begin{figure}[tp]
\centering
\label{architecture}
\includegraphics[width=3.5in]{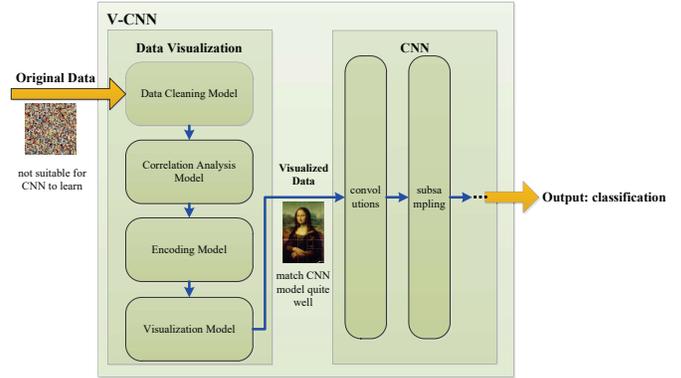}
\caption{Architecture of V-CNN.}
\end{figure}

To let arbitrary datasets match the features of CNN, V-CNN is propsoed, as shown in Fig. 1. The architecture of V-CNN contains two modules: data visualization module and CNN module. The role of CNN module is the same as the traditional CNN model comprised of multiple convolutional layers and pooling layers. Thus, in this poster we mainly focus on the data visualization module. The objective of data visualization module is to transform the original data that is not suitable for CNN to visualized data that matches CNN quite well by a series of submodules or processes, including data cleaning module, correlation analysis module, encoding model, and visualization model.

Data cleaning module guarantees the data validity by removing invalid records and attributes, and recovering specific attributes of fragmentary records.

Correlation analysis module calculates the correlation between different attributes, which is quite important since the visualization module utilizes these results to re-construct the dataset.

Encoding model tries to change the value of each attribute into the range of $[0,255]$ which indicates to the RGB value of image. Dataset always possesses two types of values: categorical value and numeric value. We adopt One-Hot encoding to transform categorical values into a series of binary values. After that, we normalize all the values within $[0,255]$.

Visualization model re-constructs the dataset. One crucial principle is that let attributes with strong correlation be next to each other, while attributes with weak correlation be away from each other. Specifically, we place each numeric value into any one RGB channel of a new pixel. For categorical value, we place multiple dimensions of each categorical value into different RGB channel of a pixel or into different RGB channels of adjacent pixels. After that, the original dataset is transformed to the visualized dataset.

%\Blindtext

\section{Use Case for Network Intrusion Detection and results}
We highlight that V-CNN is a general methodology that can be widely used in many fields. In this poster, we show one use case by applying V-CNN into network intrusion detection field. We use the AWID dataset \cite{AWID}, which is a practical dataset of WiFi network by capturing packets from realtime wireless environment. This dataset covers four kinds of network intrusion: normal, injection, impersonation, and flooding. Each record has 155 attributes. The training dataset has 1,795,575 records while the test dataset has 575,643 records. Fig. 2 presents the re-constructed visualized data in image way.

It is quite interesting that the visualized data is really something like an image. The classification quality measured with recall rate is shown in Tab. 1. It can be observed that V-CNN significantly outperforms other related studies. And, we believe that the network intrusion problem has almost been solved, at least for the AWID dataset, since recall rate of each invasion category is more than 99.8\%.

\begin{figure}[tp]
\centering
\subfigure[Normal.]{\includegraphics{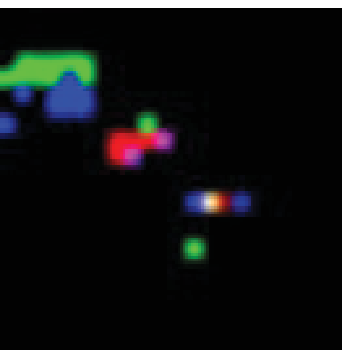}}
\label{fig:2-1}
\hfil
\subfigure[Injection.]{\includegraphics{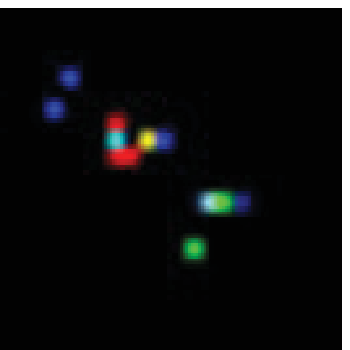}}
\label{fig:2-1}
\hfil
\subfigure[Impersonation.]{\includegraphics{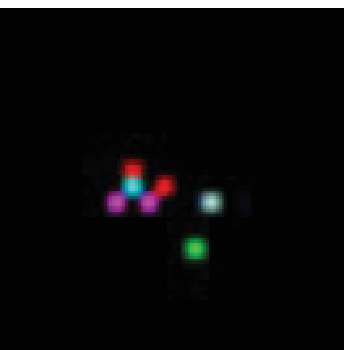}}
\label{fig:2-3}
\hfil
\subfigure[Flooding.]{\includegraphics{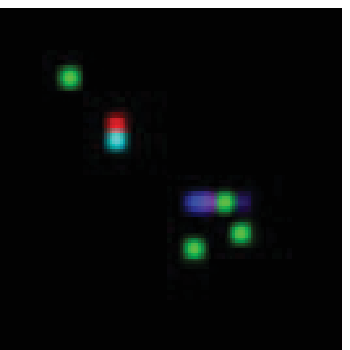}}
\label{fig:2-4}
\hfil
\caption{Data after visualization.}
\end{figure}

\section{Conclusion}
%\blindtext
Introducing data visualizing for CNN, this poster proposes a general methodology named V-CNN. V-CNN transform original arbitrary dataset that is not suitable for CNN to visualized data that matches CNN quite well. We apply V-CNN into network intrusion detection scenario, and the recall rate is more than 99.8\%, which significantly outperforms the related studies. More importantly, we highlight that V-CNN is a quite general methodology that can be widely used in many fields.

\section*{Acknowledgment}

This work was supported in part by the National Natural Science Foundations of CHINA (Grant No. 61501373, No. 61771390, No. 61771392, and No. 61271279), the National Science and Technology Major Project (Grant No. 2016ZX03001018-004), and the Fundamental Research Funds for the Central Universities (Grant No. 3102017ZY018).

\bibliographystyle{ACM-Reference-Format}
\bibliography{reference}

\end{document}